\documentclass[aps,prl,twocolumn,superscriptaddress,showpacs,amsmath,amssymb,longbibliography]{revtex4-1}
\usepackage[english]{babel}
\usepackage{amsmath}
\usepackage{bm}
\usepackage{graphicx,bbm} 
\usepackage[T1]{fontenc}
\usepackage{newtxtext,newtxmath}
\usepackage{epsfig} 
\usepackage{soul}
\usepackage[utf8]{inputenc}
\usepackage[colorlinks,linkcolor=blue,citecolor=blue,urlcolor=blue]{hyperref}
\usepackage{color}
\usepackage{latexsym}
\usepackage{ulem}
\definecolor{nred} {RGB}{224,0,0}
\definecolor{nblue} {RGB}{28,130,185}
\definecolor{dgreen} {RGB}{78,138,21}
\definecolor{norange}{RGB}{230,120,20}

\begin{document} 
\title{Spin subdiffusion in  perturbed infinite-$U$ Hubbard chain}
\author{J. R{\k{e}}kas}
\affiliation{Department of Theoretical Physics, Faculty of Fundamental Problems of Technology, Wroc\l aw University of Science and Technology, 50-370 Wroc\l aw, Poland}
\author{M. Mierzejewski}
\affiliation{Department of Theoretical Physics, Faculty of Fundamental Problems of Technology, Wroc\l aw University of Science and Technology, 50-370 Wroc\l aw, Poland}
\author{Z. Lenar\v{c}i\v{c}}
\affiliation{Jo\v zef Stefan Institute, SI-1000 Ljubljana, Slovenia}
\author{P. Prelov\v{s}ek}
\affiliation{Jo\v zef Stefan Institute, SI-1000 Ljubljana, Slovenia}

\date{\today}
\begin{abstract}
The $t$-model represents the Hubbard model in the limit $U \to \infty$ and is one of the basic models of strongly correlated electrons. On a one-dimensional chain, the model is integrable, and the charge dynamics corresponds to that of free spinless fermions. However, the sequence of spins is frozen, leading to the Hilbert space fragmentation and nontrivial spin dynamics. We consider integrable and perturbed models with perturbations that break integrability while preserving fragmentation, and show that they exhibit various types of spin dynamics, from ballistic transport to anomalous diffusion in the integrable case, and from diffusion to subdiffusion in the perturbed case. Due to fragmentation, in all cases considered, spin transport is mediated by charge transport, with a particular magnetization dependence, most notably leading to subdiffusion in the grandcanonical average of the perturbed model, with a mechanism distinct from subdiffusion in disordered or dipole-conserving models.


\end{abstract}

\maketitle


\noindent {\it Introduction.} 
The transport in integrable quantum lattice models has been
in  the focus of numerous  theoretical studies with, both, analytical and numerical 
approaches. The vast majority dealt with (in general anisotropic) Heisenberg
XXZ spin chain $1/2$ \cite{bertini21}. Another example is the 
one-dimensional (1D)  Hubbard model \cite{lieb68,ogata90}, which is even
closer to most challenging questions of materials with correlated electrons, such as
cuprates \cite{dagotto94} and fermionic cold-atom systems \cite{bordia16}, but much less investigated 
theoretically with respect to transport \cite{zotos97,prelovsek04,ilievski17,ilievski18}. 
The complexity of the 1D Hubbard model is strongly reduced going into the regime $U \gg t$ 
where it maps to the $t$-$J$ model \cite{chao78} and finally in the $U \to \infty$ limit into the integrable 
$t$-model, which has been analytically considered
in a few studies \cite{ogata90,bertini17,tartaglia22}.  Owing to the analogy with the much more  studied easy-axis XXZ model
\cite{znidaric11,medenjak17,gopalakrishnan19}, and in particular with the folded
XXZ model  at $\Delta \to \infty$   \cite{zadnik21}, the spin excitations in the 
$t$-model at zero magnetization are expected to spread via anomalous (dissipationless) diffusion.
\cite{prelovsek04,steinigeweg12,ilievski18,denardis18,prelovsek22}.

Similarly to the folded XXZ model \cite{zadnik21}, the $t$-model is an example of the 
system with the Hilbert-space fragmentation \cite{rakovszky20}.  Dynamically decoupled subspaces 
correspond to (and can be labeled by) distinct  
spin sequences because particles are impenetrable. 
The fragmentation persists within a perturbed nonintegrable 
$t$-model, where
the hopping is spin-dependent, $t_\uparrow \neq t_\downarrow$, or if one omits in the $t-J$ model the spin exchange part $J_\perp =0$, leading to 
the $t-J_z$ model \cite{batista00,rakovszky20}.  In spite
of the  integrability-breaking perturbation, the 
spin dynamics and transport are nontrivial because of  the Hilbert-space fragmentation. 
We show that such perturbed models  
offers the manifestation of different transport scenarios for spin excitations. 

Away from half filling, the basic integrable $t$-model shows  
ballistic transport at finite magnetization. However, in the canonical sector with magnetization $m=0$, there is no spin current possible even in systems with periodic 
boundary conditions (PBC). We show this both analytically and numerically by Thouless  argument of sensitivity to magnetic flux (twisted boundary conditions)  \cite{edwards72,pawlowski25}.
Similarly, we show that boundary Lindblad spin-flipping terms also cannot  induce a steady-state (dc) spin current. However, the thermodynamic grandcanonical (GC) averaging over all sectors reveals for the integrable $t$-model a dissipationless spin diffusion, similar to that in the gapped regime of the XXZ model \cite{prelovsek04,steinigeweg12,ilievski18,denardis18,prelovsek22}.

Like in the XXZ models \cite{denardis22,nandy23}, breaking of integrability in the perturbed model reduces the 
spin transport at finite magnetization from ballistic to  diffusive. Even more interesting is the GC average of transport over all sectors, where our results suggest a subdiffusive spin transport. A similar behavior has been previously theoretically anticipated in Ref. \cite{denardis22} for perturbed folded XXZ model, besides a more phenomenological treatment in Ref.~\cite{McCulloch26}, to our knowledge the only other study suggesting subdiffusion in microscopic systems without either the dipole-moment conservation \cite{gromov20,feldmeier20,sanchez20,nandy24} or disorder \cite{agarwal15,lenarcic20}. The mechanism for subdiffusion in our study differs from the latter two and can be explained in terms of solution to the porous medium equation.

\noindent{\it Perturbed $t$-model}.
We study the generalized 1D $t$-model, which can be considered as the $U \to \infty$
limit of the Hubbard model,
\begin{equation}
H_t =- \sum_{l,\sigma=\uparrow,\downarrow}
t_\sigma
\tilde c^\dagger_{l+1,\sigma} \tilde c_{l,\sigma} 
+\mathrm{H.c.} \;.\label{tjz}
\end{equation}
Double occupation of sites is not allowed and, consequently, we use projected 
fermionic operators $ \tilde c_{i,\sigma} = P_i c_{i,\sigma} P_i$ with  $ P_i = 1-n_{i \uparrow} n_{i\downarrow}$ and $n_{i \sigma}=c^{\dagger}_{i,\sigma} c_{i,\sigma}$.
Although the model  is integrable for $t_\uparrow = t_\downarrow$, as is also the parent Hubbard model \cite{ogata90}, we will also consider the case of $t-\Delta t$ model with  $t_{\uparrow,\downarrow}=t\pm \Delta t/2$, which is nonintegrable for $\Delta t \ne 0$. Another
generalization is the $t-J_z$ model, where spin interaction (without exchange) is added
\begin{equation}
H_z= H_t+ J_z \sum_i S^z_i S^z_{i+1},\qquad S^z_i =\frac{1} {2}(n_{i\uparrow} - n_{i\downarrow}).
\label{tjz2}
\end{equation}
The charge and spin currents are given, respectively, by
\begin{eqnarray}
j_c &= & \sum_{l}
j_{l,l+1,\uparrow}+j_{l,l+1,\downarrow}\;\;, \\
j_s &= & \frac{1}{2}\sum_{l}
j_{l,l+1,\uparrow}-j_{l,l+1,\downarrow}\;\;,
\end{eqnarray}
where $j_{l,l',\sigma} = (i t_{\sigma} \tilde c^\dagger_{l',\sigma} \tilde c_{l,\sigma} + \mathrm{H.c.}) $ are the local current operators.

\noindent{\it Charge current}.
First, we discuss  the  integrable  $t$-model with $\Delta t =0$, where the charge current is a conserved quantity (CQ), as 
observed long ago \cite{brinkman70}. It is interesting to note  that this CQ emerges from conserved energy current in the Hubbard model, \cite{shastry86,grabowski95,zotos97},
\begin{equation}
Q_3 = t \sum_{l,\sigma} [ j_{l-1,l+1,\sigma} - U  (n_{l,-\sigma} - \frac{1}{2})( j_{l,l+1,\sigma} +  j_{l-1,l,\sigma}) ]\;. \label{q3hub}
\end{equation}
In the case of strong interaction, $U/t \gg 1$, the first terms can be neglected along with the contribution $n_{l,-\sigma}(j_{l,l+1,\sigma} +  j_{l-1,l,\sigma})$ that requires a double occupancy of the site $l$. 
This yields $Q_3/U=j_c$, which implies the conservation of $j_c$ for $U\to \infty$. 
Consequently, the charge transport within the $t$-model is ballistic \cite{ogata90,bertini17,tartaglia22}.  Perturbations 
$\Delta t \neq 0$ or $J_z \neq 0$ break integrability so that $j_c$ is not conserved any more, leading to scattering and  diffusive dc transport in the resulting fermionic models.  

On the other hand,  the spin dynamics in the $t$-model is nontrivial. While the positions of spin-up and spin-down fermions evolve in time, the sequence of spins remains frozen (up to translations, for PBC), giving rise to Hilbert-space fragmentation \cite{rakovszky20}.  The latter  originates from the absence of  doubly occupied sites and from the fact that  hopping is restricted to neighboring sites. Consequently, fragmentation persists also for $\Delta t \ne 0$ and $J_z \ne 0$ which break the integrability along with the conservation of $j_c$.  

\noindent{\it Spin current}.
We study a chain  of $L$ sites with PBC containing $N = N_\uparrow + N_\downarrow$ particles with total spin \mbox{$S^z_{tot}= (N_\uparrow -N_\downarrow)/2$}.  First, we show that for zero magnetization,
$ m = 2 S^z_{tot}/L =0$, all diagonal matrix elements  of spin current vanish, i.e.,  $\langle n| j_s |n \rangle =0 $, where
$H_t|n\rangle=E_n |n\rangle$. To this end, we discuss the 
sensitivity to spin-dependent flux \cite{edwards72,pawlowski25}. That is, we replace \mbox{$t_{\uparrow}  \to t_{\uparrow} \mathrm{e}^{ i \varphi}$} and
\mbox{$t_{\downarrow}  \to t_{\downarrow} \mathrm{e}^{- i \varphi}$} and note that the spin current equals $j_s=  -\frac{1}{2} {\rm d} H_t/{\rm d} \varphi$ and then $\langle n| j_s |n \rangle
= -\frac{1}{2} {\rm d} E_n /{\rm d} \varphi$. 
Appropriate gauge transformation shifts the $\varphi$-dependence of $t_{\sigma}$ to an arbitrarily single bond, e.g., to the bond between site 1 and site $L$. For this reason, the derivative ${\rm d} E_n /{\rm d} \varphi$, can  be viewed as the Thouless level sensitivity (LS) to the  boundary conditions.  
The specific feature of the present LS is that the corresponding boundary conditions are spin-dependent, so the LS contains information on the spin transport instead of the charge transport.  In the following, we distinguish between the results obtained within a single magnetization sector, $m$, from GC averaging over various magnetization sectors. For the latter, we explicitly indicate the average magnetization, $\langle m \rangle_{GC}$.

\begin{figure}[tb]
\includegraphics[width=1.0\columnwidth]{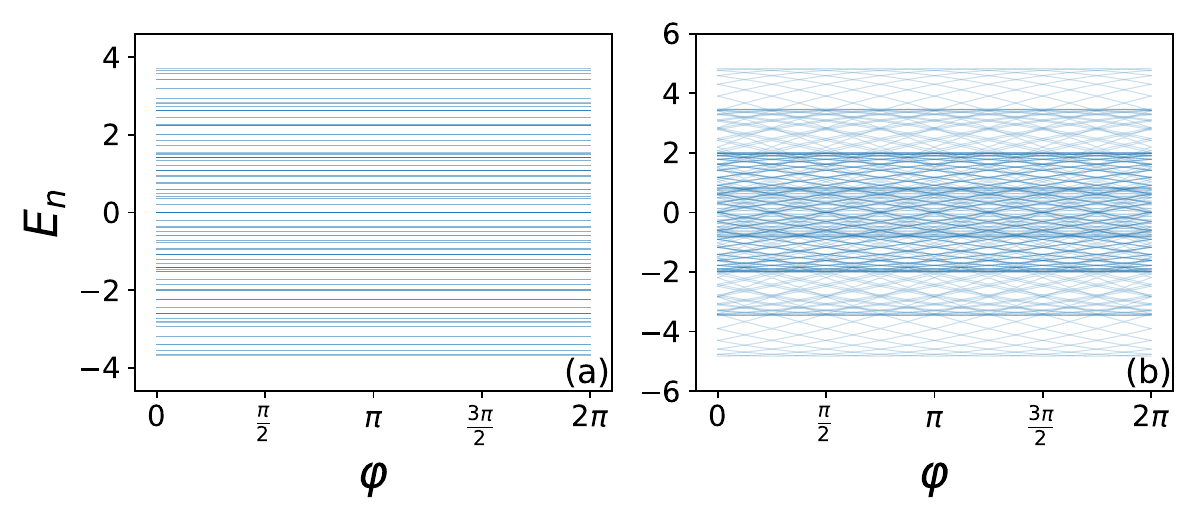}
\caption{ Spectrum of the integrable $t$-model 
vs. flux $\varphi$ for a system of $L =8$ sites. 
(a) shows results for magnetization $m =0$  with 
$N_\uparrow =N_\downarrow = 3$, and (b) for $m \ne 0 $
with $N_\uparrow =2 $ and $N_\downarrow =3$.} 
\label{fig1}
\end{figure}

Fig.\ref{fig1} shows the energy levels, $E_n(\varphi)$, in a small system   described by the $t$-model.  
For  $m=0$, the LS vanishes as shown in Fig.\ref{fig1}(a).  It is evident from Fig.~\ref{fig1}(b) that $m \neq 0$ is required to obtain the nonvanishing flux dependence of $E_n(\varphi)$ and consequently also  $\langle n| j_s |n \rangle \neq 0$. The same conclusions 
hold  for other system sizes and particle densities, as well 
as for $\Delta t \ne 0$ and $J_z \ne 0$ (not shown). In the
Supplemental Material (SM) \cite{sm}, we present 
analytical arguments for the vanishing of LS at $m=0$.

To gain  physical insight into the LS results, one may consider a time-dependent flux with a sufficiently small slope, ${\rm d \varphi}/{\rm d t}={\rm const.}$, such that the system remains in the linear-response regime. 
This flux drives spin-up particles to move along the chain in one direction and spin-down particles in the opposite direction. At $m=0$, the charge current must vanish due to symmetry, while the vanishing of the LS also implies the absence of the steady spin current.  For nonzero $m$, the same flux induces a charge current flowing in the direction of the particles with the majority spin polarization and nonvanishing LS indicates a finite spin current. Consequently, the LS results shown in Fig.~\ref{fig1} strongly suggest that spin and charge currents in the $t$-model are intrinsically coupled to one another. In the following, we show this intrinsic coupling via explicit calculations for an open system.   

{\it Open systems}. 
As an alternative to the time-dependent driving, we consider an open system with edge coupling to the baths  
\begin{equation}
    \frac{\partial{\rho}}{\partial \tau}=-i[{H}_{t(z)},{\rho}]+\sum_{k=1}^2 {L}_k{\rho}{L}_k^\dagger -\frac{1}{2}( {L}_k^\dagger {L}_k{\rho}+{\rho}{L}_k^\dagger{L}_k).
\label{eq:lindblad}
\end{equation}
We use edge dissipators ${L}_1={S}_1^+=\tilde{c}_{1\uparrow}^\dagger\tilde{c}_{1\downarrow}$ and
${L}_2={S}_L^-$
that can induce spin current without causing any imbalance in the charge sector.
There is a class of initial states \mbox{$\rho_i=|\uparrow,\vec{s},\downarrow,\alpha\rangle
\langle \uparrow,\vec{s},\downarrow,\alpha |$} that are annihilated by the dissipator and thus immune to current generation via coupling to baths at any time. These are projectors on eigenstates of the Hamiltonian $H_{t}$ or $H_{z}$ where  the spin of the first particle is up spin and that of the last particle is down. In the above notation, $\vec{s}=(\sigma_2,\ldots,\sigma_{N-1} )$ labels the spin configuration of other $N-2$ fermions in the system, while $\alpha$ enumerates eigenstates within a fragmented subspace. Since $\rho_i$ are the steady states of Eq.~(\ref{eq:lindblad}) as well as of closed systems (described by $H_{t}$ or $H_{z}$) with open boundary conditions, they do not support any steady current. 

\begin{figure}[tb]
\includegraphics[angle =0, width=0.9\columnwidth]{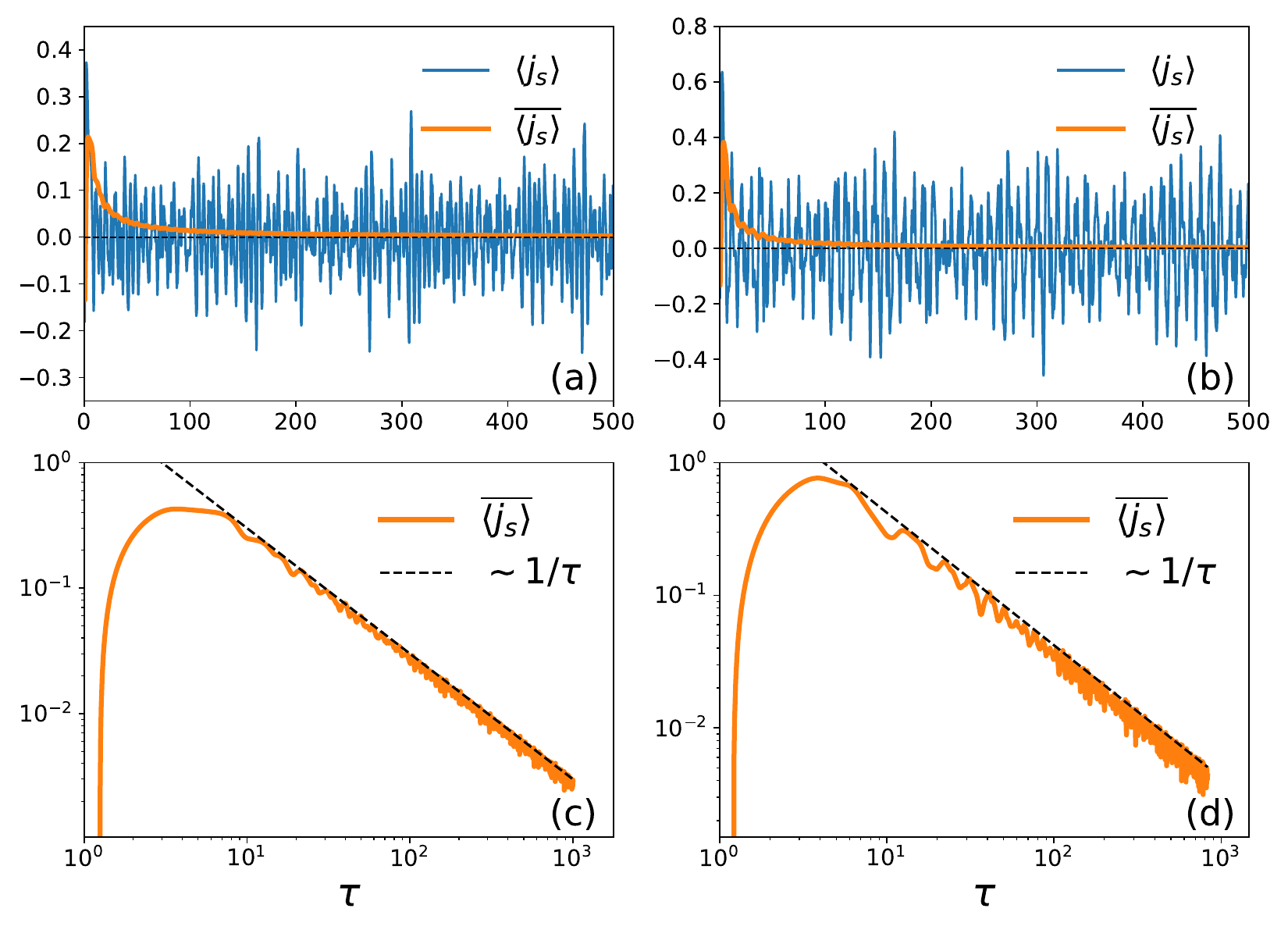}
\caption{(a),(b) Time-dependence of the spin current, $\langle j_s \rangle(\tau)$, obtained from numerical solution of Eq.\eqref{eq:lindblad} for $H_t$, given by Eq.\eqref{tjz} with $\Delta t=0$ on $L=8$. We chose the initial state
$|\downarrow0\uparrow0\downarrow0\uparrow0\rangle$ in (a) and $|\downarrow0\uparrow0\uparrow0\uparrow0\rangle$ in (b). (c) and (d) show cumulative averages of the spin currents shown in (a) and (b), respectively. 
} 
\label{fig2n}
\end{figure}

For other initial states, we  determine the spin current from the numerical solution of Eq.~(\ref{eq:lindblad}), $\langle j_s \rangle(\tau)={\rm Tr} [\rho(\tau) j_s] $. The results in Figs.~\ref{fig2n}(a,b) show strong undamped oscillations of the spin current for two different initial states. In SM \cite{sm} we link these oscillations to a strongly structured spectrum of the Liouvillian (\ref{eq:lindblad}) containing multiple non-decaying states (with zero real and finite imaginary part of eigenvalue) with finite current expectation value. In Figs.~\ref{fig2n}(c,d) we plot the cumulative averages $\overline{\langle j_s \rangle}(\tau)=\frac{1}{\tau}\int_0^\tau {\rm d\tau'}  \langle j_s \rangle(\tau')$ that decay in time
 as $\tau^{-1}$, indicating the absence of  a stationary spin current. This is also consistent with spectrum analysis, which reveals that steady states with zero eigenvalue has zero spin current expectation value. It implies absence of the dc spin current independent of the initial state, expected also on increasingly large system sizes unless one includes Lindblad operators that allow for dc charge current, as shown in \cite{sm}. To summarize this part we note that that charge and spin currents are inherently coupled. Systems that do not allow for a dc charge current likewise do not support a dc spin current.

{\it Anomalous spin diffusion in the integrable $t$-model.}
From now on, we consider closed systems with $\varphi=0$ and PBC, which allow for a dc charge current. The nontrivial spin transport in the $t$-model then  emerges from the overlap with $j_c$.  The relevant overlaps  can be evaluated analytically at $T \to \infty$ and $L \to \infty$,
\begin{equation}
\langle j_c j_s \rangle = -2 L m (1-n) \overline t^2, \quad  \langle j_c j_c \rangle = 2 L n (1-n) \overline t^2,
\end{equation}
where $\langle ... \rangle={\rm Tr}()/{\rm Tr}(1) $, 
$\overline t^2 = (t^2_\uparrow + t^2_\downarrow)/2 = t^2+ \Delta t^2/4$, 
and $n=N/L$  is the particle density.   
In the integrable model, $j_c$ is conserved, leading to a finite spin stiffness $D_s$ in each sector with $m \neq 0$ according to the Mazur-based inequality \cite{zotos97},
\begin{equation}
T D_s \sim \frac{1}{2L} \frac{ \langle j_c j_s \rangle^2}{ \langle j_c j_c \rangle} =
\frac{ m^2 (1-n) t^2 } {n}.
\label{mazur}
\end{equation}
The same $m^2$-dependence of $D_s$  occurs within the easy-axis XXZ model \cite{medenjak17,ilievski18,prelovsek22}.
In the GC ensemble at $\langle m \rangle_{GC}  =0$ such $m^2$-dependence leads in the thermodynamic limit  to a finite spin diffusion constant, ${\cal D}^0_s$, even though the transport remains dissipationless. The diffusive  transport originates from the boundary scattering with  the effective relaxation time \mbox{$\tau_L \propto L$}
and the grand canonical average, $ \langle m^2 \rangle_{GC} \propto 1/L $ that combined together give an $L$-independent dc diffusion ${\cal D}^0_s \propto \tau_L \langle D_s \rangle_{GC} $.

The dynamical spin diffusion can be evaluated within the linear-response theory, ${\cal D}_s(\omega) = \int_0^\infty d \tau \mathrm{e}^{i \omega \tau}  
\langle j_s( \tau) j_s \rangle/(L \tilde \chi_s)$, where 
$T \chi_s = \tilde \chi_s = n/4$ is the (renormalized) spin susceptibility at $T \to \infty$.
In the following we discuss results for 
${\cal D}_s(\omega)$,  obtained numerically via the microcanonical Lanczos method (MCLM) 
\cite{long03,prelovsek13,prelovsek21}. The method allows for system sizes up to $L =20$ and the frequency-resolution $\delta \omega \simeq 10^{-3}$ 
by employing $\sim 10^4$ Lanczos steps  in the present study.  

\begin{figure}[tb]
\includegraphics[width=1.0\columnwidth]{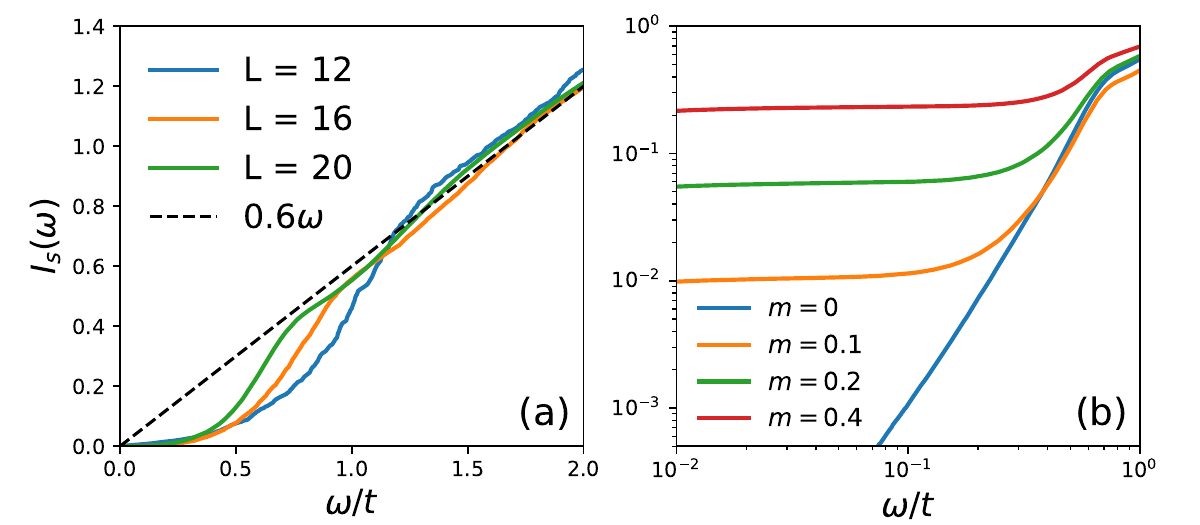}
\caption{ (a) Integrated spin diffusion $I_s(\omega)$, as calculated within the integrable $t$-model at $ n = 1/2$ for magnetization $m=0$ using MCLM for 
systems with PBC on $L = 12, 16, 20$ sites. Dotted line denotes the simplest
$L \to \infty$ extrapolation. (b) $I_s(\omega)$  
(in log-log scale) for $L = 20$ sites, but for different magnetizations  $m =0, 0.1, 0.2, 0.4$.  } 
\label{fig2}
\end{figure}

The dissipationless diffusion in integrable 
\mbox{$t$-model} can be approached in finite systems in various ways. In Fig.~\ref{fig2}(a)
we present the result for the integrated diffusion function, $I_s(\omega) = \int_0^\omega d \omega' {\cal D}_s(\omega')$, 
in the $m=0$  sector at quarter filling, i.e., at particle density $n=1/2 $. The above arguments may suggest vanishing of the spin diffusion constant,
${\cal D}^0_s={\cal D}_s(\omega \to 0) =0$ for fixed magnetization $m=0$ and finite $L$. However, it has been  recognized \cite{prelovsek04} that 
the the dc response of macroscopic systems,  $\lim_{\omega \to 0} \lim_{L\to \infty}$ is not trivial, since large finite-size oscillations converge to diffusive behavior
$I_s(\omega) \sim \omega {\cal D}^0_s $ where ${\cal D}^0_s \sim 0.6 t$. On the other hand,
results in Fig.~\ref{fig2}(b) for $m \ne 0$ clearly reveal (at fixed  $n \sim 1/2$) a finite spin stiffness $D_s >0$,
visible  as  $I_s(\omega  \to 0) \propto D_s \propto m^2$.  

\begin{figure}[t!]
\includegraphics[width=1.0\columnwidth]{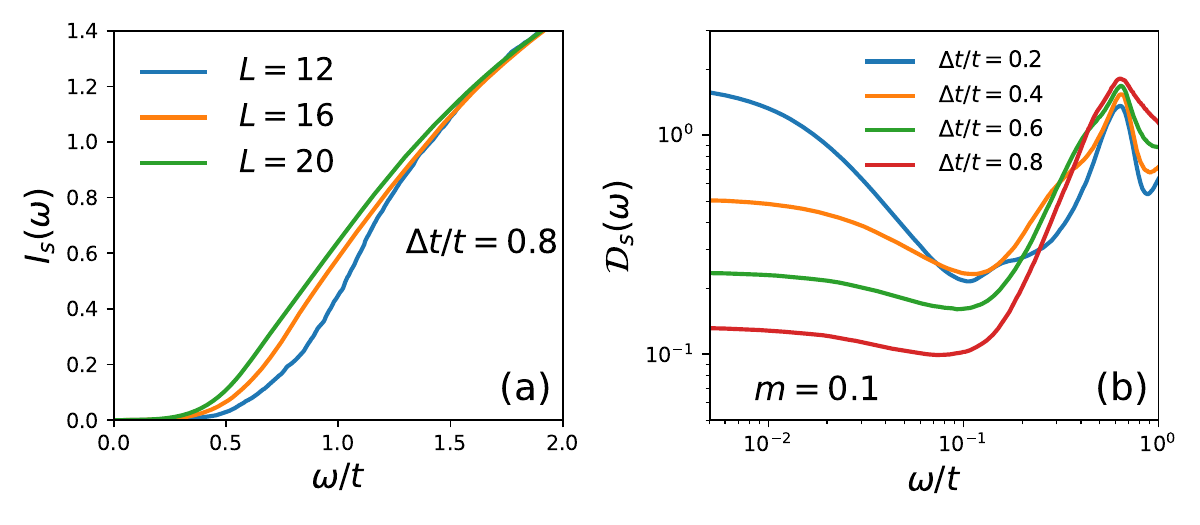}
\caption{ (a) Integrated $I_s(\omega)$, as calculated
within perturbed $t$-model with $\Delta t/t =0.8$ at $ n = 1/2$ and $S^z_{tot}=0$ 
 on $L = 12, 16, 20$ sites. (b) Dynamical ${\cal D}_s(\omega)$ (in log-log scale) calculated for $L = 20$ sites at $n = 1/2$ and $m = 0.1$, for different $\Delta t / t = 0.2 - 0.8$.} 
\label{fig3}
\end{figure}

\noindent {\it Transport in the perturbed $t$-model.}
Away from half-filling, $ n \neq 1$, the integrability breaking perturbations reduce the charge transport from ballistic  to 
normal diffusive, depending on the strength of perturbation, i.e., either ${\cal D}^0_c \propto 1/\Delta t^2$ or ${\cal D}^0_c \propto 1/J_z^2$. This is expected to hold in each sector with $m \neq 0$. Nevertheless,  the restrictions on spin transport due to fragmentation remain the same, i.e., the spin current appears through the overlap with the charge current. Thus a clear difference with respect to the integrable model should appear also for GC averaged case with
$\langle m \rangle_{GC}=0$. While
in the integrable case,  $I_s(\omega) \propto \omega$ is consistent with diffusive spin transport, as evident in Fig.~ \ref{fig2}(a), the scaling of $I_s(\omega)$ at finite perturbation $\Delta t \ne 0$ suggests vanishing dc 
${\cal D}^0_s \to 0$ even at $L \to \infty$, as presented 
in Fig.~\ref{fig3}(a) for  $\Delta t = 0.8t$. The canonical sector 
with $m=0$ differs markedly from $m\ne0 $ sectors, as 
the latter clearly exhibit finite spin diffusion consistent with overlaps with diffusive $j_c$.
The results in Fig.~\ref{fig3}(b) for $m=0.1$ at different $\Delta t/t = 0.2 - 0.8$
reveal that the perturbation broadens the spin-stiffness
$\delta(\omega)$-peak of the integrable $t$-model
to a relaxation peak with finite  $ {\cal D}^0_s \propto D_s(m)/\Delta t^2$.  Very similar 
conclusions follow from the results for $t-J_z$ model, presented in SM \cite{sm}. 
     
\noindent {\it Evidence for subdiffusion.}
In the thermodynamic limit ($L\to \infty$), the integrable $t$-model shows the same type of spin transport as the easy axis XXZ chain
\cite{prelovsek04,steinigeweg12,ilievski18,denardis18,prelovsek22}, i.e., ballistic transport at $m \neq 0$ and 
dissipationless diffusion (within the GC ensemble) at $ \langle m \rangle_{GC}  =0$. 
On the other hand, the perturbed  $t$-model
should exhibit a different scenario. At
fixed  $m \neq 0$  the spin diffusion is normal and originates from the 
broadening of the  stiffness-$\delta(\omega)$-peak, as it is for charge transport. According to Eq.~\eqref{mazur},  the weight associated with the $\delta(\omega)$-peak grows with magnetization as $m^2$, while the 
relaxation time depends on the perturbation, $\propto 1/\Delta t^2$. Then, for each
magnetization sector one expects the spin diffusion ${\cal D}^0_s\propto m^2/\Delta t^2$. Consequently, the diffusion constant relevant for spreading of the coarse grained density of magnetic excitations in hydrodynamic region $x$, $u(x)=\langle S^z_i \rangle_{i\in x}$, is proportional  to squared local magnetization, $u^2(x)$, and one obtains a  nonlinear subdiffusion equation 
\begin{equation}
\partial_{\tau} u(x,\tau) \propto
\partial_x \left(u(x,\tau)^2 \partial_x u(x,\tau)\right) \propto \partial_{xx}u(x,\tau)^3 \; . \label{pme}
\end{equation}
Eq. \eqref{pme} is known as the {\it porous medium equations} and it has been intensively studied in the context of nonlinear transport, see \cite{Vazquez2006} for review. In particular, the spread of the density profile $d(x,0)\propto \delta(x)$ is described by the Barenblatt solution $u(x,\tau)=\tau^{-1/4}f(|x|\tau^{-1/4})$.     

\begin{figure}[tb]
\includegraphics[width=0.7\columnwidth]{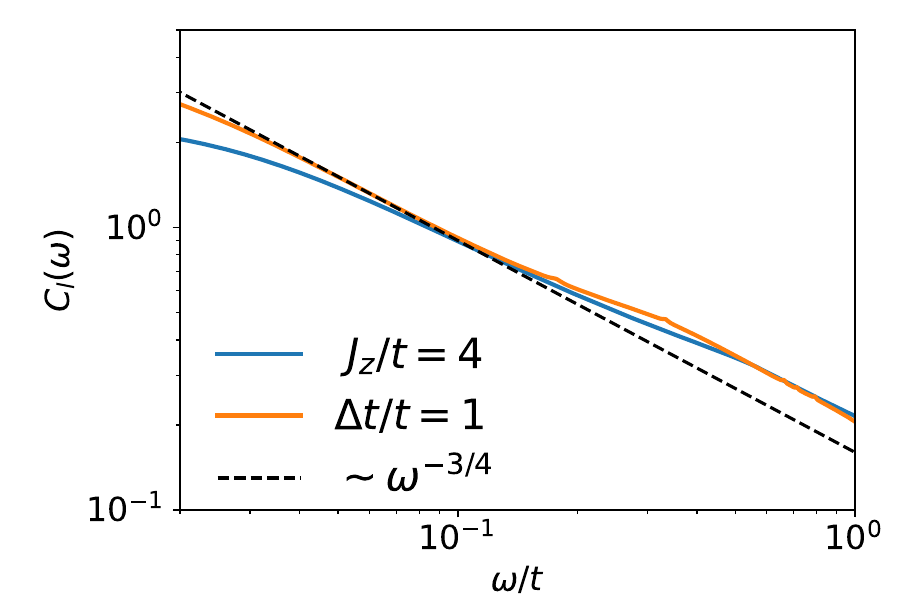}
\caption{ Local spin correlations $C_l(\omega)$ (in the log-log scale),
at $n=1/2$  and $m=0$  calculated on $L = 20$ sites within the $t-J_z$ model at $J_z/t = 4$,  as well as within  $t-\Delta t$ at $\Delta t/t = 1$~. The dotted
line represents the power-law dependence 
$\propto \omega^{-3/4}$.  }
\label{fig4}
\end{figure}

In finite systems, obtaining direct numerical evidence for subdiffusion is challenging. It requires sufficiently large 
$L$ and long times for the relevant scattering processes to dominate over finite-size effects. In the following, we calculate the local  spin correlation function, $C_l(\omega) =   \int_0^\infty \mathrm{e}^{i \omega \tau}  
\langle S^z_l(\tau) S^z_l \rangle  d\tau $  at $m =0$, 
in a system with PBC. 
Based on Eq.~\eqref{pme}, one expects
the relation
$C_l(\omega)\propto \int_0^\infty \mathrm{e}^{i \omega \tau} u(0,\tau)\propto \omega^{-\gamma}$ with the exponent
$\gamma=1-1/4=3/4$. In finite-size systems the scaling should be visible   for intermediate times $\tau$ or,
equivalently, for  intermediate frequencies $\omega$, provided that 
perturbation is large enough. 
In Fig.~\ref{fig4} we present $C_l(\omega)$ both for $t-J_z$ model (with $J_z/t=4$)
and for $t-\Delta t$ model ($\Delta t/t=1$) which both are compatible with 
the exponent $\gamma \simeq 3/4$.

Ref.~\cite{denardis22} previously gave an argument for subdiffusion in the perturbed folded XXZ model. While the subdiffusive transport in the $\Delta\to\infty$ limit of XXZ model should be mediated by diffusion of magnons through domains of varying but fixed size, the origin of subdiffusive transport in perturbed $t$-model could be related to diffusive transport of holes through subregions of spins configurations pointing in the same direction, with lengths of such polarized regions varying randomly.

\noindent{\it Allowing the spin exchange.} 
It is evident that any perturbation which would break the Hilbert-space 
fragmentation by allowing change of spin sequence, would lead also to
normal diffusion at $m=0$. In the SM \cite{sm} we show this explicitly by including the spin exchange via $J_\perp$ term.

\noindent{\it Conclusions.} We have studied spin transport 
in the integrable and generalized $t$-model, considering perturbations that break the integrability of the parent Hamiltonian while preserving the  Hilbert-space fragmentation. 
We find that in the thermodynamic limit or within the GC ensemble, the integrable $t$-model 
shows the dissipationless diffusive spin transport while the perturbed (but fragmented) $t$-models exhibit subdiffusive spin-transport, a very unique phenomenon 
in translationally invariant models, recently claimed also for the folded XXZ model \cite{denardis22}.

Hilbert space fragmentation manifests itself through a frozen spin sequence and leads to a crucial role of the coupling between charge and spin currents. We have demonstrated this numerically via the (Thouless-type) level sensitivity to spin-dependent flux 
in a system with PBC and through calculations of the spin current in a boundary-driven open system. Both  approaches show that within a fixed magnetization sector $m =0$ spin current strictly vanishes. This is the case even at $m \ne 0$ in a open system, if boundary-driving is solely by spin exchange. Nevertheless, the inclusion of a boundary term permitting charge exchange enables a finite spin current.

The coupling between the spin and charge currents increases with magnetization as $m^2$. Within the integrable 
$t$-model, a diffusive yet nondissipative spin current emerges 
in the GC ensemble as a consequence of ballistic charge transport 
combined with the averaging over magnetization sectors, 
as is the case also within the easy-axis XXZ chain \cite{prelovsek04,steinigeweg12,ilievski18,denardis18,prelovsek22}. 
For the perturbed but fragmented $t$-model, charge transport becomes diffusive, with diffusion constant dependent on the magnetization as $m^2$, yielding subdiffusive spin transport in the $L \to \infty$ limit or within the GC ensemble, confirmed also by explicit calculations of local correlation functions. 
We can conclude tat spin transport in the 
$t$-model exhibits many similarities to that of the folded XXZ model \cite{zadnik21}, while offering a more intuitive and physical picture due to its origin in the infinite-$U$ Hubbard model \cite{lieb68,ogata90}.

\begin{acknowledgments}
The authors thank L. Zadnik, S. Nandy and J. De Nardis for fruitful discussions. Z.L. and P.P. acknowledge the support by the 
project N1-0088 and program P1-0044  of the Slovenian Research and Innovation Agency (ARIS), and the ERC StG 2022 project DrumS by Horizon Europe, Grant Agreement 101077265, and the European Union Horizon 2020 under the QuantERA II project QuSiED (No 101017733).

\end{acknowledgments}

\bibliography{bibliography
}
\newpage
\phantom{a}
\newpage
\setcounter{figure}{0}
\setcounter{equation}{0}
\setcounter{page}{0}

\renewcommand{\thetable}{S\arabic{table}}
\renewcommand{\thefigure}{S\arabic{figure}}
\renewcommand{\theequation}{S\arabic{equation}}
\renewcommand{\thepage}{S\arabic{page}}

\renewcommand{\thesection}{S\arabic{section}}

\onecolumngrid

\begin{center}
{\large \bf Supplemental Material:\\
Spin subdiffusion in  perturbed infinite-$U$ Hubbard chain}\\
\vspace{0.3cm}
J. R\k{e}kas$^{1}$, M. Mierzejewski$^{1}$, Z. Lenar\v{c}i\v{c}$^{2}$ and P. Prelov\v{s}ek$^{2}$,\\
$^1${\it Department of Theoretical Physics, Faculty of Fundamental Problems of Technology, \\ Wroc\l aw University of Science and Technology, 50-370 Wroc\l aw, Poland}\\
$^2${\it Department of Theoretical Physics, J. Stefan Institute, SI-1000 Ljubljana, Slovenia} \\
\end{center}

In the Supplemental Material we present in more detail several aspects: the analytical arguments for the vanishing spin current at zero magnetization, numerical results for the spin diffusion in the 
$t$-$J_z$ model as well for the systems where the spin exchange is allowed.  We discuss also the transport in open systems.

\vspace{0.6cm}

\twocolumngrid

\label{pagesupp}

\section{Level sensitivity in the generalized $t$-model} \label{app1}

The key argument for the vanishing of the spin current at $S^z_{tot} =0$
follows from the fact that in integrable as well as perturbed (i.e. with
$J_z , \Delta t \neq 0$) $t$-model particles are impenetrable and 
consequently distinguishable. Hence,
they can be considered to emerge from (one) squeezed wf. with 
$N = N_\uparrow + N_\downarrow$ particles on $L$ sites,
\begin{equation}
| \phi^{\underline s}_i \rangle = c^\dagger_{i+N,s_N}c^\dagger_{i+N-1,s_{N-1}}\cdots  c^\dagger_{i,s_1} | 0 \rangle,
\end{equation}
which has a particular sequence of spins $\underline s$. The general basis wf. can be obtained by unsqueezing  $| \phi^{\underline s}_i \rangle$ (but still with fixed initial $i$ and $s_1$),
\begin{equation}
| \phi^{\underline s}_{i,\underline d} \rangle = c^\dagger_{i+N +d_N,s_N} \cdots  c^\dagger_{i+1+d_1,s_2} c^\dagger_{i,s_1} | 0 \rangle,
\end{equation}
where the sequence of shifts $\underline d$ should satisfy $d_N \geq d_{N-1} \geq \cdots d_2 \geq 0$. Then, general
eigenfunction in the case of PBC can be searched in the form
\begin{equation}
| \Psi^{\underline s}_q  \rangle = \sum _{l, \underline d} \mathrm{e}^{iql} \alpha_{\underline d} ~
T_l  | \phi^{\underline s}_{i,\underline d} \rangle,
\qquad T_l   | \phi^{\underline s}_{i,\underline d} \rangle =  | \phi^{\underline s}_{i +l,\underline d} \rangle, \label{phiq}
\end{equation}

We are now looking for possible diagonal matrix elements of Hamiltonian (to any power) $H_t^k, k \geq 1$ within basis wf. and eigenstates, Eq.~(\ref{phiq}). 
Such matrix elements $H^k$, e.g., appear in diagonalization via 
Lanczos procedure or in the Hamiltonian time evolution.
The point is that any elements which would
show flux dependence would also imply the corresponding
eigenvalue dependence $E(\varphi)$, which indicates (via Thouless approach \cite{edwards72}) in a system with PRB nonvanishing spin current. Let us consider  the lowest-order matrix element of $H^N_t$ between wf. shifted by a single site (in analogy also for shifts by several sites),
\begin{equation}
\langle \phi^{\underline s}_{i,\underline d} | H_t^N | \phi^{\underline s}_{i+1,\underline d} \rangle =
t_\uparrow^{N_\uparrow} t_\downarrow^{N_\downarrow} \mathrm{e}^{i  \varphi (N_\uparrow-N_\downarrow)} 
\propto  \mathrm{e}^{2 i  \varphi S^z_{tot} }.\label{matr}
\end{equation}
For such a matrix element it is enough to show that at $S^z_{tot} \neq 0$ 
there is in general nonvanishing $\varphi$ dependence, while for $S^z_{tot} =0$ Eq.~(\ref{matr}) is $\varphi$-independent. 

In the case $S^z_{tot} =0$ evidently there can be nontrivial elements
between wf. with different shift sequences $\underline d$. However, 
one can for the case $N_\uparrow = N_\downarrow$ redefine 
basis function Eq.~(\ref{phiq}), including the flux $\varphi$,
\begin{eqnarray}
| \Psi^{\underline s}_q  \rangle &=& \sum _{l, \underline d} \mathrm{e}^{iql} \tilde \alpha_{\underline d} ~
T_l  | \tilde \phi^{\underline s}_{i,\underline d} \rangle,
\quad | \tilde \phi^{\underline s}_{i,\underline d} \rangle =  
\mathrm{e}^{i \eta_{\underline d} \varphi} | \phi^{\underline s}_{i,\underline d} \rangle, \nonumber \\
\eta_{\underline d} &=& \sum_{j\uparrow =1}^{N_{\uparrow}}d_{j \uparrow} - \sum_{j\downarrow =1}^{N_{\downarrow}} d_{j \downarrow}.
\end{eqnarray}  
There are in general different-order matrix elements of Hamiltonian $H^k$,
with $k = k_\uparrow + k_\downarrow$ which connect such states.
But a general observation (for redefined basis functions) in
the case $N_\uparrow = N_\downarrow$ is that matrix elements 
remains independent of $\varphi$, i.e.,
\begin{equation}
\langle \tilde \phi^{\underline s}_{i,\underline d'} |H_t^k | \tilde \phi^{\underline s}_{i,\underline d} \rangle 
= t_\uparrow^{k_\uparrow} t_\downarrow^{k_\downarrow}.
\end{equation}
This leads to the conclusion that in case of $S^z_{tot} =0$ 
all considered matrix elements are independent of $\varphi$.
Moreover, this fact we confirm also by direct numerical evaluation of dependence $E_n(\varphi)$ in finite systems (see main text),
where we show that at $S^z_{tot} =0$ levels $E_n(\varphi) = E_n$ 
are independent of $\varphi$. Consequently, this leads to
vanishing diagonal spin current $j_{s,nn} \propto d E_n/d \varphi = 0$,
while in general this is not the case for $S^z_{tot} \neq 0$.

\section{Spectrum of the
Liouvillian in an open system} 
\label{app2}
The Liouville operator
\begin{equation}
    \hat{\mathcal{L}}=-i[ {H}, {\rho}]+\sum_{k=1}^2\left(  {L}_k {\rho} {L}_k^\dagger -\frac{1}{2}(  {L}_k^\dagger {L}_k {\rho}+ {\rho} {L}_k^\dagger {L}_k)\right)
\label{eq:lindbladSM}
\end{equation}
with spin-flipping edge Lindblad terms
\begin{equation}
 {L}_1= {S}_1^+= \tilde{c}_{1\uparrow}^\dagger \tilde{c}_{1\downarrow}, \quad  {L}_2= {S}_L^-= \tilde{c}_{L\downarrow}^\dagger \tilde{c}_{L\uparrow}
\label{eq:dyssy}
\end{equation}
is considered in two complementary ways. 
For the results shown in the main text, we obtain the time dependent spin current expectation value $\langle j_s \rangle(\tau)={\rm Tr} [\rho(\tau) j_s] $ by performing the time evolution of some initial density matrix $e^{\hat{\mathcal{L}} \tau}\rho_i$ using Runge-Kutta method  of fourth order with time step $dt=10^{-2}$ in $L=8$ sites. As commented earlier, we observe strong oscillations with cumulative average $\overline{\langle j_s \rangle}(\tau)=\frac{1}{\tau}\int_0^{\tau} {\rm d \tau'}  \langle j_s \rangle(\tau') \propto \tau^{-1} $ that vanishes in time.

\begin{figure}[t!]
\includegraphics[width=0.8\columnwidth]{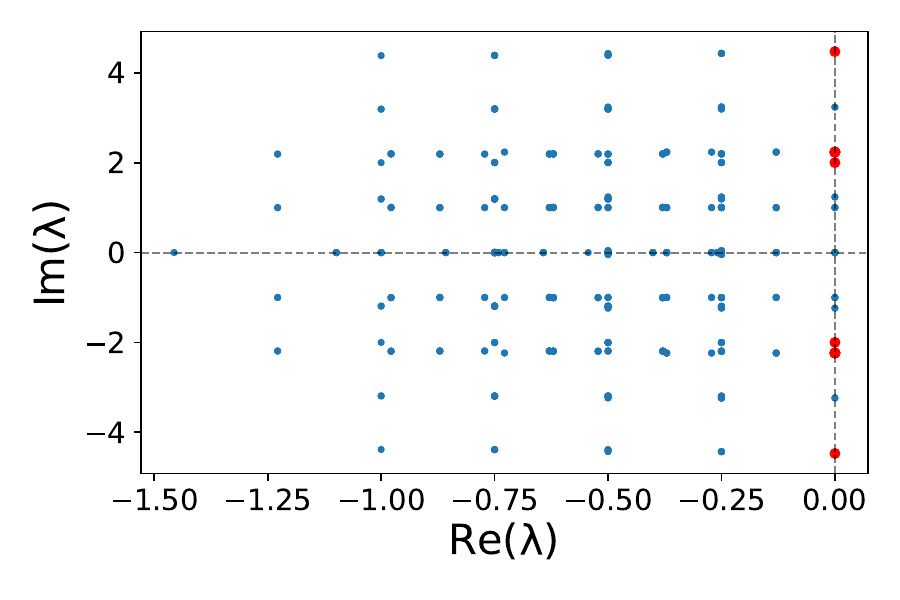}
\caption{Liouville spectrum calculated for Hamilotnian $H_t$, see Eq.\eqref{tjz} in the main text, with $L=4$ and $N=2$ fermions. Red dots correspond to eigenvectors with non-zero spin current expectation value and zero real part of eigenvalue.} 
\label{figggg5}
\end{figure}
\begin{figure}[b!]
\includegraphics[width=0.8\columnwidth]{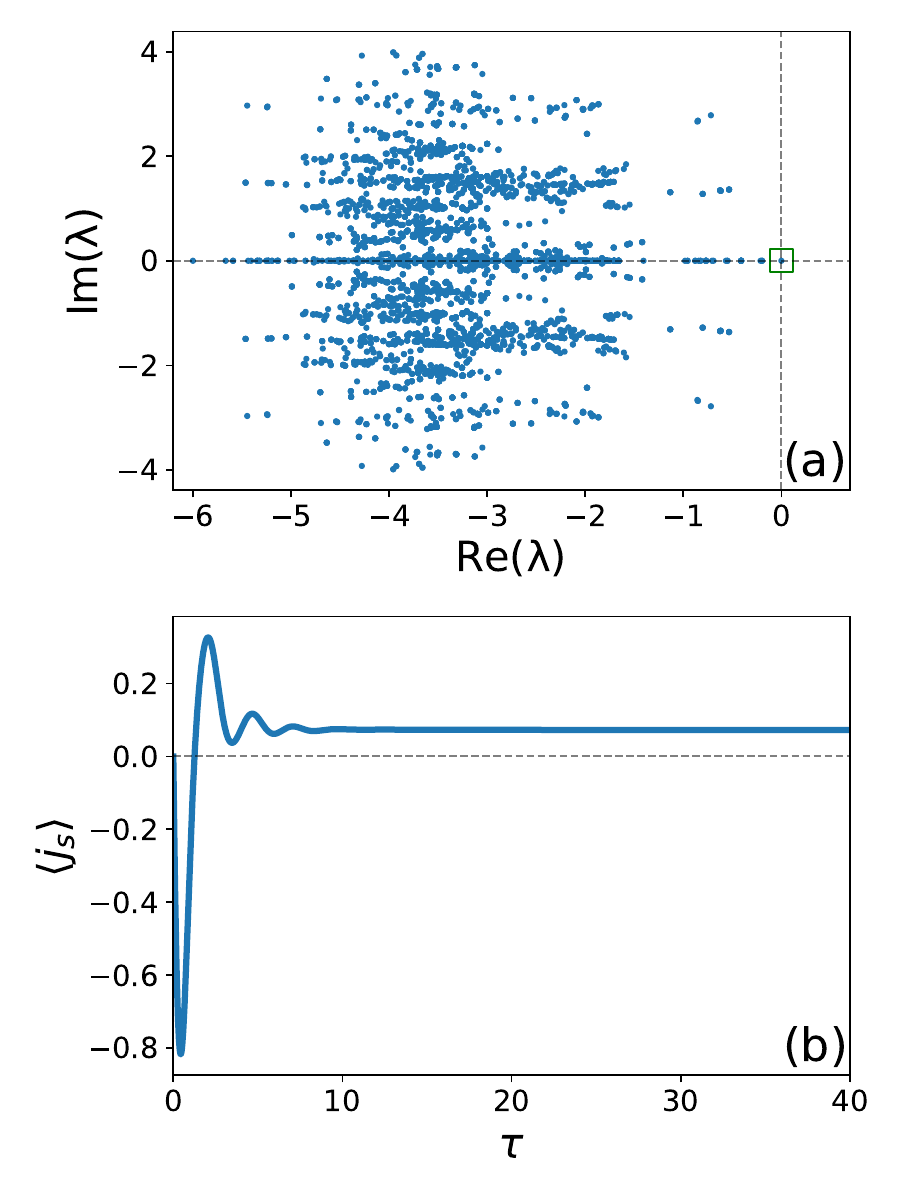}
\caption{(a) Liouville spectrum calculated on $L=4$ sites for Hamiltonian $H_t$ and Lindblad operators \eqref{eq:dyssy1} in addition to previously considered \eqref{eq:dyssy}. (b) The corresponding spin current calculated on $L=6$ with the initial state $|\downarrow 0\uparrow\downarrow 0 \uparrow \rangle$. } 
\label{js:all-L}
\end{figure}

Such oscillating behavior with zero steady state value can be explained with the properties of the Liouville spectrum, shown in Fig.~\ref{figggg5}. Spectrum is calculated on smaller system with $L=4$ sites. An extremely regular structure of the spectrum is revealed as a consequence of the fragmentation. Importantly, there exist several states $\rho_m$ with zero real and finite imaginary part of the eigenvalue. Non-decaying states with finite expectation value of current ${\rm Tr} [j_s \rho_m]$ are on Fig.~\ref{figggg5} denoted by red points; they are the cause of non-decaying oscillation in the time dependent spin current expectation value. The steady state $\rho_\infty$ with zero real and imaginary part of the eigenvalue, on the other hand, does not carry current ${\rm Tr} [j_s \rho_{\infty} ]=0$, consistent with zero steady state spin current.

In the following, we demonstrate that the anomalous dynamics shown in Fig.~\ref{fig2n}, as well as the highly specific Liouville spectrum in Fig.~\ref{figggg5}, disappear once charge fluctuations are allowed. To this end, in addition to the Lindblad terms from Eq.~\eqref{eq:dyssy}, we introduce  also other Lindblad operators, $L_3,\dots,L_{10}$, 
\begin{equation}
L_{3,4} =\tilde{c}_{1\uparrow,\downarrow}\;, \quad L_{5,6} =\tilde{c}_{L\uparrow, \downarrow}\;, \quad 
L_{7,8,9,10}=L^{\dagger}_{3,4,5,6}\;.
\label{eq:dyssy1}
\end{equation}
The latter dissipators by themselves do not generate any bias voltage that could drive the charge transport. Nevertheless, they allow a charge to flow through the studied open system.  

We repeated the calculation shown in Fig.~\ref{fig2n} of the main text using the enlarged set of dissipators. The resulting spin current, displayed in Fig.~\ref{js:all-L}, exhibits a steady-state value reached after a relatively short transient dynamics.  The Liouville spectrum obtained with the full set of dissipators 
$L_1,\dots,L_{10}$ is also shown in Fig.~\ref{js:all-L}. Unlike the case shown in Fig. \ref{figggg5}, the spectrum now exhibits a unique eigenstate with eigenvalue having zero real part. This state is stationary, as its imaginary part equals zero as well.

\section{Spin transport within the $t-J_z$ model}  

\begin{figure}[b!]
\includegraphics[width=1.0\columnwidth]{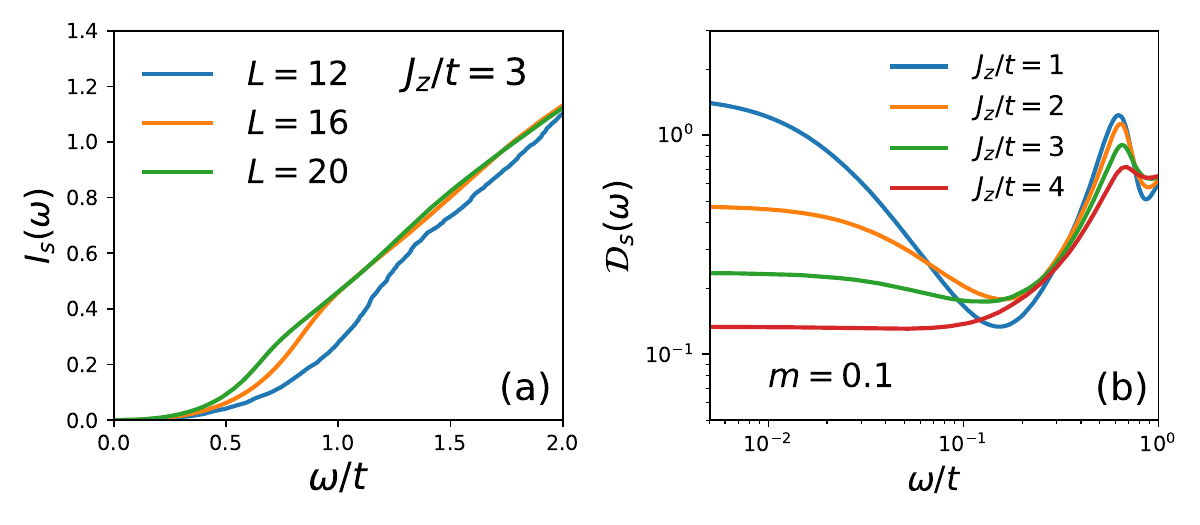}
\caption{(a) Integrated spin diffusion $I_s(\omega)$, as calculated
within $t - J_z$ model, Eq.~\eqref{tjz2} in the main text, for $J_z/t = 3$ at $ n = 1/2$ 
and zero magnetization $m=0$ on chains with $L = 12, 16, 20$ sites. (b) Spin diffusion ${\cal D}_s(\omega)$ (log-log plot) calculated for $L = 20$ sites at $n = 1/2$ and $m = 0.1$ for different $J_z / t = 1 - 4$.} 
\label{fig5}
\end{figure}

While in the main text we concentrate on results with
$\Delta t \neq 0$ perturbation, we consider here analogous
results for the $J_z \neq 0$ perturbation, i.e.,
we study the model from Eq.~\eqref{tjz2} in the main text.
In Fig.~\ref{fig5} we present results for different 
$J_z$, analogous to Fig.~\ref{fig3} with different 
$\Delta t$. The integrated diffusion $I_s(\omega)$ 
in Fig.~\ref{fig5}(a) calculated at zero magnetization, 
$m=0$, and quarter filling, $n=1/2$, at fixed 
$J_z/t =3$ (note that quite  large $J_z$ is needed to 
induce significant perturbation),  but within different sizes 
$L = 12-20$. It confirms that, in contrast to integrable
$J_z=0$ in Fig.~\ref{fig2}, the limit $L \to \infty$ 
would be consistent  with vanishing dc value  
${\cal D}_s^0 =0 $. On the other hand, at nonzero magnetization, 
$m =0.1$, in  Fig~\ref{fig5}(b) the dynamical diffusion functions
${\cal D}_s(\omega)$ reveal finite dc value ${\cal D}_s^0$
decreasing with $J_z$, approximately following
perturbative result ${\cal D}_s^0 \propto 1/J_z^2$.

\section{ Breaking fragmentation by the spin exchange}  

\begin{figure}[tb]
\includegraphics[width=0.7\columnwidth]{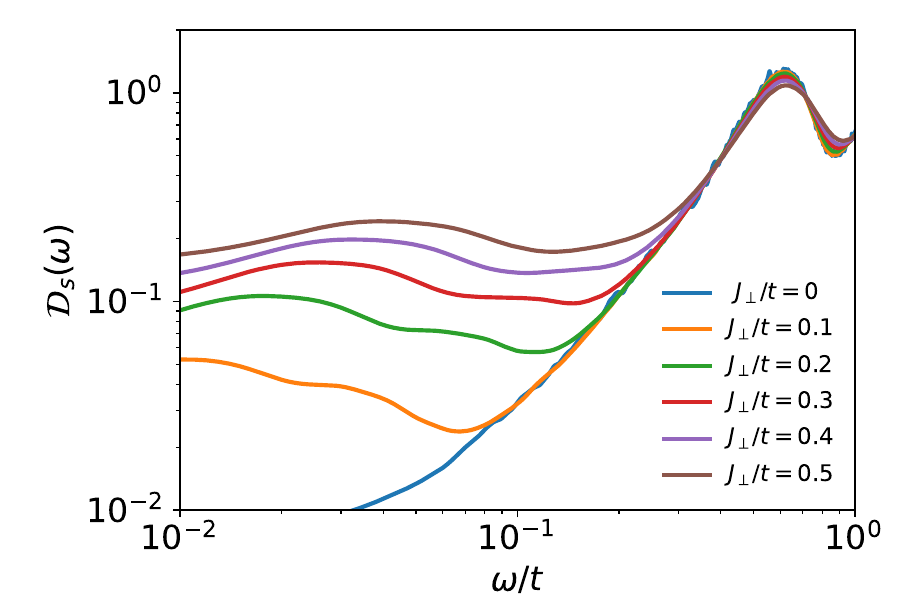}
\caption{ Dynamical spin diffusion ${\cal D}_s(\omega)$ 
(log-log plot), as calculated at $n=1/2$  and $m =0$  
on $L = 20$ sites within the 
model from Eq.~\eqref{tjz3} at different $J_\perp/t = 0 - 0.4$ and for $J_z=0$.}
\label{fig6}
\end{figure}

One expects that any additional perturbation that enables rearrangement of the spin sequence, thus breaking Hilbert-space fragmentation, results in a qualitatively different transport scenario. Such cases
can be realized, e.g.,  by adding to the $t-J_z$ model, Eq.~\eqref{tjz2} in the main text, the spin exchange  term, 
\begin{equation}
H_\perp= H_z+\frac{1}{2} J_\perp \sum_i (S^+_{i+1} S^-_i + S^-_{i+1} S^+_i),
\label{tjz3}
\end{equation}
which for $J_\perp = J_z = J$ correspond to the full $t - J$ model \cite{ogata90}. 
Another possibility is to allow for the next-neighbor hopping. 
In general, such cases would lead (at $T \to \infty$) to diffusive spin and charge transport    even within the canonical  $m=0$
sector. We present in Fig.~\ref{fig6} the corresponding result for
${\cal D}_s(\omega)$ at quarter filling $n =1/2$ and $m=0$ calculated on $L=20$ system for the model from Eq.~\eqref{tjz3} with $J_z=0$.
Various curves show results for different $J_\perp/t = 0 - 0.4$.
For simplicity, we neglected 
 the contribution to the spin current, $j_s$, which are of the order \mbox{of  $\sim J_{\perp}$}. It is evident that $J_\perp \neq 0$
induces even at $m =0$ finite dc diffusion ${\cal D}^0_s >0$ with the transport becoming close to results within 
the 1D $t-J$ model \cite{bonca95}, i.e. for $J_z=J_\perp =J$. 

\end{document}